\documentclass{article}

\usepackage{xcolor}
\usepackage{hyperref}
\usepackage{graphicx}
\usepackage{authblk}
\usepackage{lipsum}
\usepackage[backend=biber,style=numeric]{biblatex}
\usepackage{glossaries}
\usepackage{cleveref}
\usepackage{subcaption}
\title{Enhancing Cell Counting through MLOps: A Structured Approach for Automated Cell Analysis}

\author[1,2,3*]{Matteo Testi}
\author[4,*]{Luca Clissa}
\author[5,*]{Matteo Ballabio}
\author[1,2*]{Salvatore Ricciardi}
\author[4]{Federico Baldo}
\author[6]{Emanuele Frontoni}
\author[7]{Sara Moccia}
\author[8]{Gennaro Vessio}

\affil[1]{Deep Learning Italia, Milan, Italy}
\affil[2]{Artificial Intelligence Venture Builder, London, UK}
\affil[3]{Università Campus Biomedico di Roma, Rome, Italy}
\affil[4]{Università di Bologna, Bologna, Italy}
\affil[5]{Università di Bergamo, Bergamo, Italy}

\affil[6]{Università di Macerata, Macerata, Italy}
\affil[7]{Department of Innovative Technologies in Medicine and Dentistry, Università degli Studi G. d'Annunzio Chieti – Pescara, Chieti, Italy}
\affil[8]{Università degli Studi di Bari Aldo Moro, Bari, Italy}

\date{}

\begin{document}

\maketitle

\def\thefootnote{*}\footnotetext[1]{These authors contributed equally to this work.}\def\thefootnote{\arabic{footnote}}

\begin{abstract}


Machine Learning (ML) models offer significant potential for advancing cell counting applications in neuroscience, medical research, pharmaceutical development, and environmental monitoring. However, implementing these models effectively requires robust operational frameworks. This paper introduces Cell Counting Machine Learning Operations (CC-MLOps), a comprehensive framework that streamlines the integration of ML in cell counting workflows. CC-MLOps encompasses data access and preprocessing, model training, monitoring, explainability features, and sustainability considerations. Through a practical use case, we demonstrate how MLOps principles can enhance model reliability, reduce human error, and enable scalable Cell Counting solutions. This work provides actionable guidance for researchers and laboratory professionals seeking to implement machine learning (ML)- powered cell counting systems.

\end{abstract}

\section{Introduction}

Cell counting is a fundamental task in biological and medical research, involving the quantification of cells in a biological specimen. These applications typically rely on collecting vast numbers of specimens, which are then observed and analyzed through microscopy imaging techniques by human operators. These operators aim to recognize and quantify the biological structures present in the samples. However, this process requires skilled researchers, is highly time-consuming, and prone to errors due to operator fatigue and subjectivity in recognizing borderline cases. These challenges represent significant bottlenecks to accelerating research and innovation in neuroscience, diagnostics, and biotechnology. As such, cell counting is an excellent candidate for automation through machine learning (ML) models, which can bring substantial benefits in terms of speed, accuracy, and reproducibility \cite{grishaginautomatcellcounting, singh2024systematic, bhattarai2024deep}.

Therefore, integrating machine learning (ML) models into cell counting applications is a promising frontier for advancing medical research, pharmaceutical development, and environmental monitoring. However, the adoption of ML models in these critical applications faces significant obstacles. First, measuring model performance is challenging due to the inherent subjectivity of the task. Second, building trust among stakeholders, such as researchers and healthcare professionals, is challenging due to the black-box nature of ML models and their lack of explainability for predictions. 
For these reasons, Machine Learning Operations (MLOps) workflows offer a promising solution by enabling solid development, clear performance assessment, reproducible results, and a robust analysis cycle, all of which are essential for sensitive applications. Despite their potential benefits, MLOps methodologies remain largely unexplored in this domain.

This paper introduces the Cell Counting MLOps (CC-MLOps) framework as a structured approach to implementing MLOps principles in cell counting applications. The primary contribution of this work is the development and validation of a comprehensive MLOps workflow optimized for cellular quantification tasks. The CC-MLOps framework encompasses multiple integrated components: data access and preprocessing protocols, model training methodologies, systematic monitoring mechanisms, explainability features, and sustainability considerations. 

The implementation of CC-MLOps principles enables significant operational improvements, including workflow optimization, error minimization through automation, enhanced model reliability via continuous validation, and infrastructure scalability. Through rigorous implementation with fluorescent microscopy images of neuronal cells during torpor studies, this framework demonstrates how MLOps best practice effectively address critical challenges in biomedical image analysis, particularly regarding reproducibility and trustworthiness. The integration of these principles into cell counting workflows represents a methodological advancement with potential applications across multiple high-risk domains. The broader adoption of such structured MLOps approaches in healthcare and environmental monitoring contexts could substantially enhance operational efficiency and analytical reliability while simultaneously facilitating continuous methodological innovation in quantitative cellular analysis.

\subsection{Related Work}

Recent research highlights the application of MLOps in various sectors, emphasizing their role in enhancing operational efficiency and scalability in ML projects. These applications range from cloud-based solutions \cite{janapareddicloud} and open-source AI architectures in agriculture\cite{cobparroagronomy} to digital twin models \cite{lombardotwins}. This diversity demonstrates the adaptability of MLOps in scaling across different domains. 
However, only a few studies have attempted to establish comprehensive MLOps guidelines in sensitive fields like neuroscience, healthcare and bioinformatics. Notably, Khattak et al. (2023) introduced a framework that addresses the complexities of machine learning (ML) in healthcare, emphasizing data governance, secure model deployment, and monitoring strategies that comply with healthcare regulations.

Most research in this field focuses on developing a standard ML pipeline, for cell counting task\cite{aicellcounter}\cite{mizuno}\cite{molder}, including data preprocessing, model evaluation, and selection to achieve accurate predictions. However, these studies often lack critical MLOps aspects such as model versioning, monitoring, and solution scaling. This study aims to fill this gap by presenting a comprehensive use case that integrates these phases of MLOps implementation, helping researchers build efficient, reliable, and scalable solutions.

\subsection{Our Contribution}
This work contributes by developing a comprehensive MLOps pipeline for examining cellular structures and counting cells in fluorescent microscopy images: CC-MLOps. Our pipeline is engineered to enhance robustness and scalability. We showcase its application in neuroscience research settings through pre-production experiments and provide a practical use case. Our study explores several critical factors essential for the successful deployment of ML models in this context:
\begin{enumerate}
    \item data access and preprocessing: we establish standardized protocols for accessing and handling data, using the \textit{yellow collection} datasets from the \textbf{Fluorescent Neuronal Cells v2} archive \cite{clissa2023fluocells} as benchmark. The preprocessing pipeline includes normalization techniques to account for intensity variations, noise reduction methods specific to fluorescence microscopy, and data augmentation strategies to enhance model generalization across diverse cellular morphologies, ensuring that the dataset used for training represents the variability encountered in real-world applications;
    
    \item ML methodology: we leverage state-of-the-art semantic segmentation architectures for cell recognition and quantification in rodent brain tissues. Using cell-ResUNet as our base architecture, we demonstrate how MLOps methodologies facilitate systematic hyperparameter optimization, focussing on loss function selection as an illustrative example of the optimization workflow that can be extended to other parameters;
    
    \item Continuous Integration and Continuous Deployment (CI/CD): we implement a CI/CD framework to automate the integration and deployment of ML models, enabling rapid iteration among interdisciplinary teams and improving cell counting performance. Our implementation integrates experiment tracking with \textit{Neptune.ai} and deployment through Google Cloud Platform's \textit{Vertex AI}, establishing a robust pipeline for model versioning and serving;
    
    \item algorithm explainability: given the importance of building trust in automated cellular analysis solutions, we demonstrate the adoption of post-hoc interpretability tools as Gradient-weighted Class Activation Mapping (Grad-CAM) visualization. These techniques provide researchers and clinicians with insights into model decision processes, facilitating validation of ML results against domain expertise and enhancing confidence in the automated cell counting workflow;
    
    \item environmental sustainability: we demonstrate how to track computational resource usage and power consumption metrics throughout the ML lifecycle. This monitoring capability addresses growing concerns about the environmental impact of ML operations, enabling researchers to assess the carbon footprint of various model configurations and training regimes. Our implementation provides a foundation for developing more sustainable machine learning (ML) solutions through resource-efficient model selection and deployment strategies.
\end{enumerate}

By thoroughly addressing these factors, our work provides deeper insights into the challenges and solutions for operationalizing ML models in cell counting for fluorescent microscopy imaging. We showcase how to apply MLOps principles in practice, sharing detailed workflow implementations and highlighting best practices that can be leveraged by similar applications — and extended to a broad spectrum of use cases — to develop effective and reliable ML solutions. The systematic approach presented here aims to set a precedent for future studies and implementations, demonstrating how MLOps principles can enhance the effectiveness and reliability of ML in clinical practice.

The rest of this paper is structured in two chapters; chapter 2 is divided in ten subsections: Problem Understanding, Data Acquisition, ML Methodology, ML Training \& Testing, Continuous integration, Continuous Delivery, Continuous Testing, Continuous Monitoring, Explainability and Sustainability while in the chapter 3 are discussed the conclusions of the work.

\section{CC-MLOps}\label{sec:cc_mlops}

\begin{figure}[t]
    \flushleft
    \includegraphics[scale=0.55]{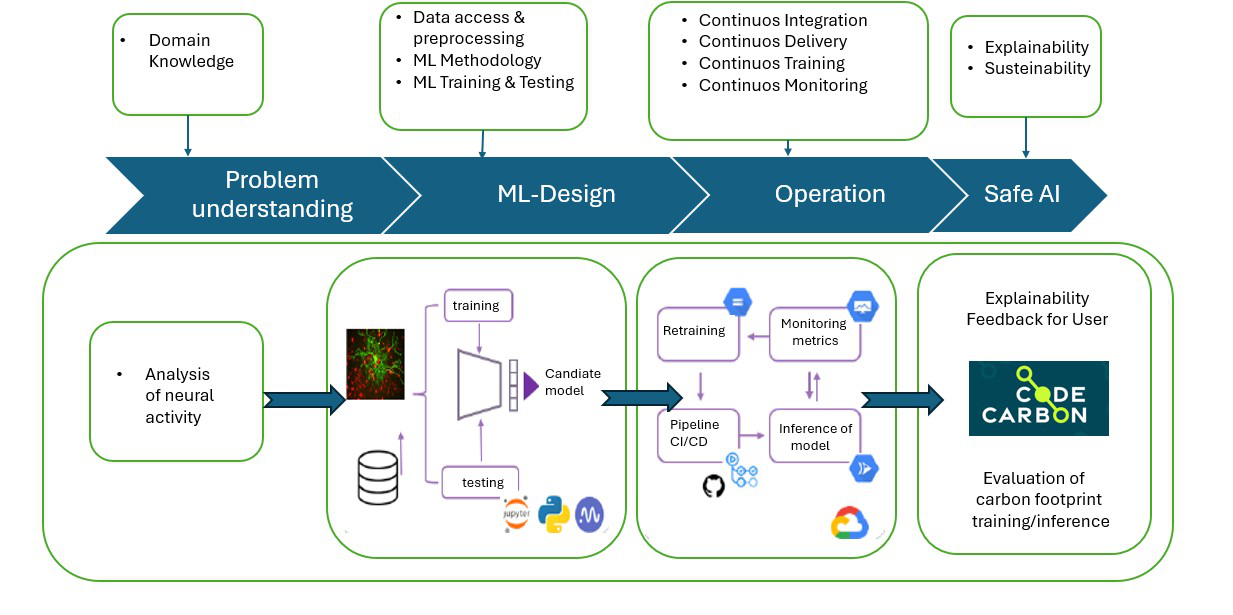}
    \caption{The CC-MLOps life-cycle. This cycle begins with understanding the problem and concludes with sustainability, encompassing solution design, integration, and deployment. It represents a continuous journey of improving and refining ML models for reliable and accurate results. }\label{fig:pipeline}
\end{figure}

In a previous work \cite{testimlops}, we presented a comprehensive methodology for implementing MLOps to establish a robust standard for machine learning (ML) projects. The well-known CRISP-DM framework is inspired this methodology \cite{crispdmprocess} and designed to bridge the gap between research and industry. Our proposed pipeline, CC-MLOps, encompasses a series of distinct steps that span the entire ML workflow, from data access and pre-processing to model deployment. Specifically, this paper focuses on the challenging task of recognizing and counting neuronal cells in fluorescence microscopy images of rodent brain tissues. Figure \ref{fig:pipeline} illustrates the various steps of our proposed pipeline, which is tailored to meet the complexities and regulatory requirements of this medical application.

In this section, we discuss the implementation details of our pipeline and highlight its adaptation for cell counting in fluorescence microscopy. We discuss the adopted methodologies and the corresponding technologies used to implement them, emphasizing the specific adjustments made to address the complexities of this application. We aim to provide a clear and detailed account of how MLOps can be effectively applied to challenging imaging tasks in sensitive applications
, ensuring robust and reproducible results.

\subsection{Problem Understanding}

Fluorescent microscopy is a vital tool in various scientific disciplines, providing a detailed view of cellular structures and functions. This technology has significantly advanced medical diagnostics and the exploration of cellular biology. Among its numerous applications, cell recognition and counting are particularly crucial in fields such as cancer research~\cite{awasthi2023fluorescence}, neurobiology~\cite{derevtsova2021applying}, and immunology~\cite{lee2020deep}, where accurate cell quantification is essential for understanding disease mechanisms and guiding therapeutic interventions.

An intriguing example of the application of fluorescent microscopy is in studies of torpor~\cite{hitrec2021reversible}, that investigate the physiological adaptations of organisms during periods of reduced metabolic activity and the brain areas involved in this process. Beyond scientific curiosity, such studies have significant implications for human applications. Understanding and reproducing the mechanisms that regulate torpor onset could allow less invasive intensive surgeries~\cite{alam2012hypothermia,bouma2012induction} and even facilitate space travel~\cite{CERRI2021cool,puspitasari2021hibernation}. Consequently, accurate cell quantification and activity analysis are critical.

Current cell recognition and counting practices often rely on human operators, highlighting the need for more efficient and automated solutions. Efforts to streamline cell counting have explored various approaches, from adaptive heuristics~\cite{luppi1,luppi2,luppi3} to deep learning methods~\cite{morelli2021cresunet,zeng2019ric_unet}. However, challenges such as domain shift~\cite{poon2023dataset}, poor generalization across diverse datasets, and a small training dataset have caused limited success~\cite{clissa2023fluocells,clissa2024scidata}. The inherent complexity of fluorescent microscopy images necessitates robust algorithms that can accurately discern subtle variations in cell morphology and staining. Additionally, deep learning solutions require careful consideration of numerous hyperparameters, including model architecture, training settings, and loss function choices, to achieve optimal performance.

Experimentation and evaluation are crucial to addressing these challenges~\cite{clissa2024iciap}. Researchers must explore various solutions, tune hyperparameters carefully, and rigorously assess performance. The complexity and numerous possible settings necessitate a systematic approach to track and compare variations in data, training pipelines, and performance metrics. Experiment-tracking tools emerge as vital components of the research workflow, enabling meticulous documentation and analysis of the impact of different approaches. These tools can significantly advance research in fluorescent microscopy and cell recognition by leveraging pre-built solutions and adapting them to specific applications.

In this work, we focus on the use case of fluorescent microscopy applied to torpor onset studies, demonstrating how MLOps tools can build a robust pipeline for model training, assessment, and deployment. Specifically, we used the \textit{yellow collection dataset} from the \textbf{Fluorescent Neuronal Cells v2} archive  \mbox{(\textbf{FNC v2 - yellow})}~\cite{clissa2024scidata,clissa2023fluocells} and the \textbf{cell-ResUNet} architecture~\cite{morelli2021cresunet} to illustrate the effectiveness of this approach in overcoming the challenges associated with automated cell counting.

\subsection{Data access and pre-processing}

The FNC v2 - yellow dataset is an open-source archive comprising 283 high-resolution, labeled fluorescent microscopy images acquired from 68 rodents under controlled experimental conditions. This dataset exhibits considerable heterogeneity in cellular characteristics, including diverse morphologies, variable sizes, and multiple orientations, as well as significant variations in brightness, luminosity, and color intensity.

These inherent variations present substantial challenges for automated cell counting systems, necessitating robust algorithms that can generalize across diverse morphological patterns and staining characteristics. The FNC v2 - yellow dataset was specifically selected for its representativeness of typical fluorescence microscopy data encountered in life science experiments, providing an appropriate test case for evaluating MLOps methodologies. For comprehensive dataset statistics, acquisition protocols and annotation methodologies, readers are directed to the original publication ~\cite{clissa2024scidata,clissa2023fluocells}.

\begin{figure}
    \centering
    \begin{subfigure}{0.45\textwidth}
        \centering
        \includegraphics[width=\linewidth]{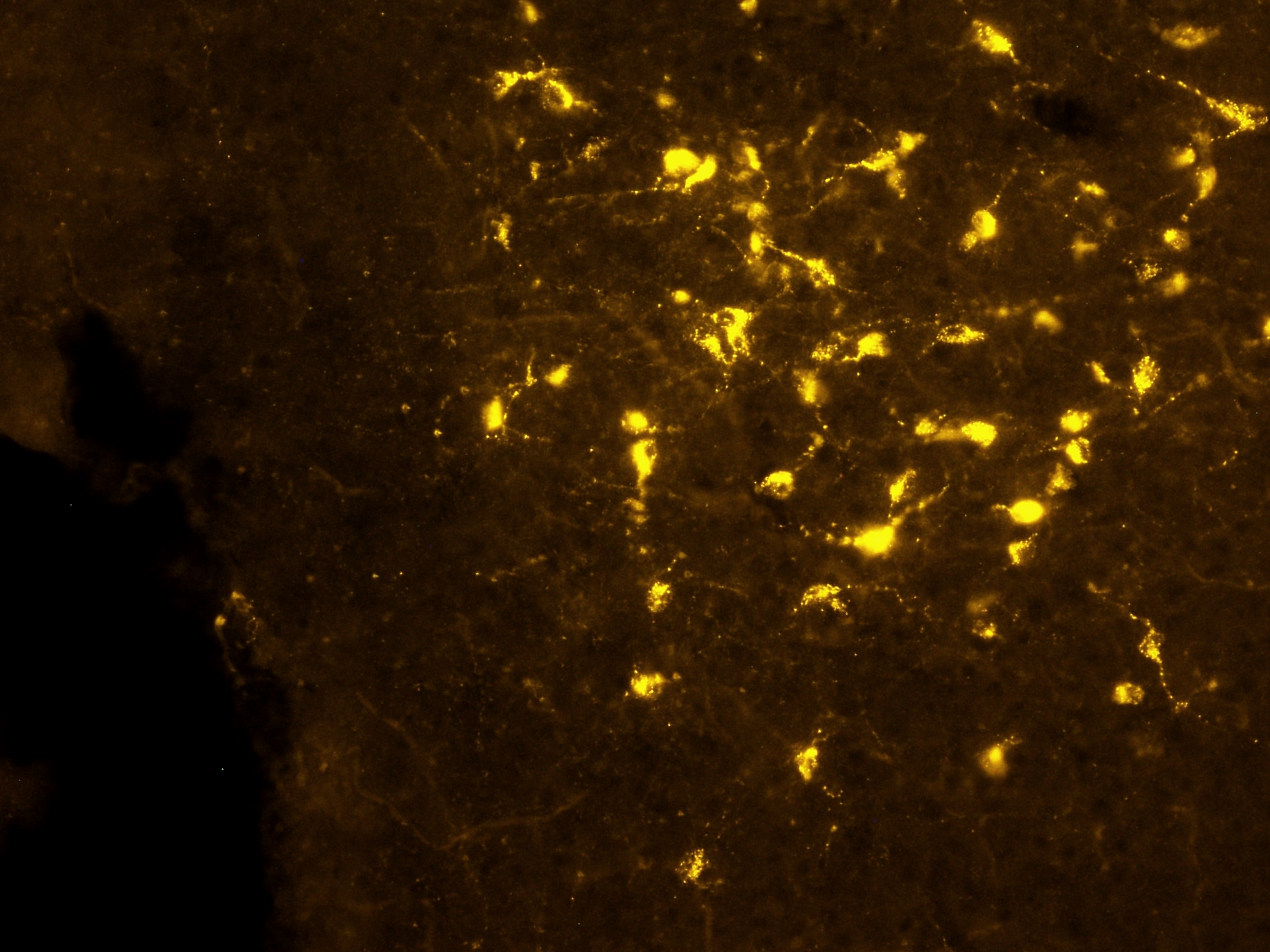}
        \caption{}
    \end{subfigure}
    \begin{subfigure}{0.45\textwidth}
        \centering
        \includegraphics[width=\linewidth]{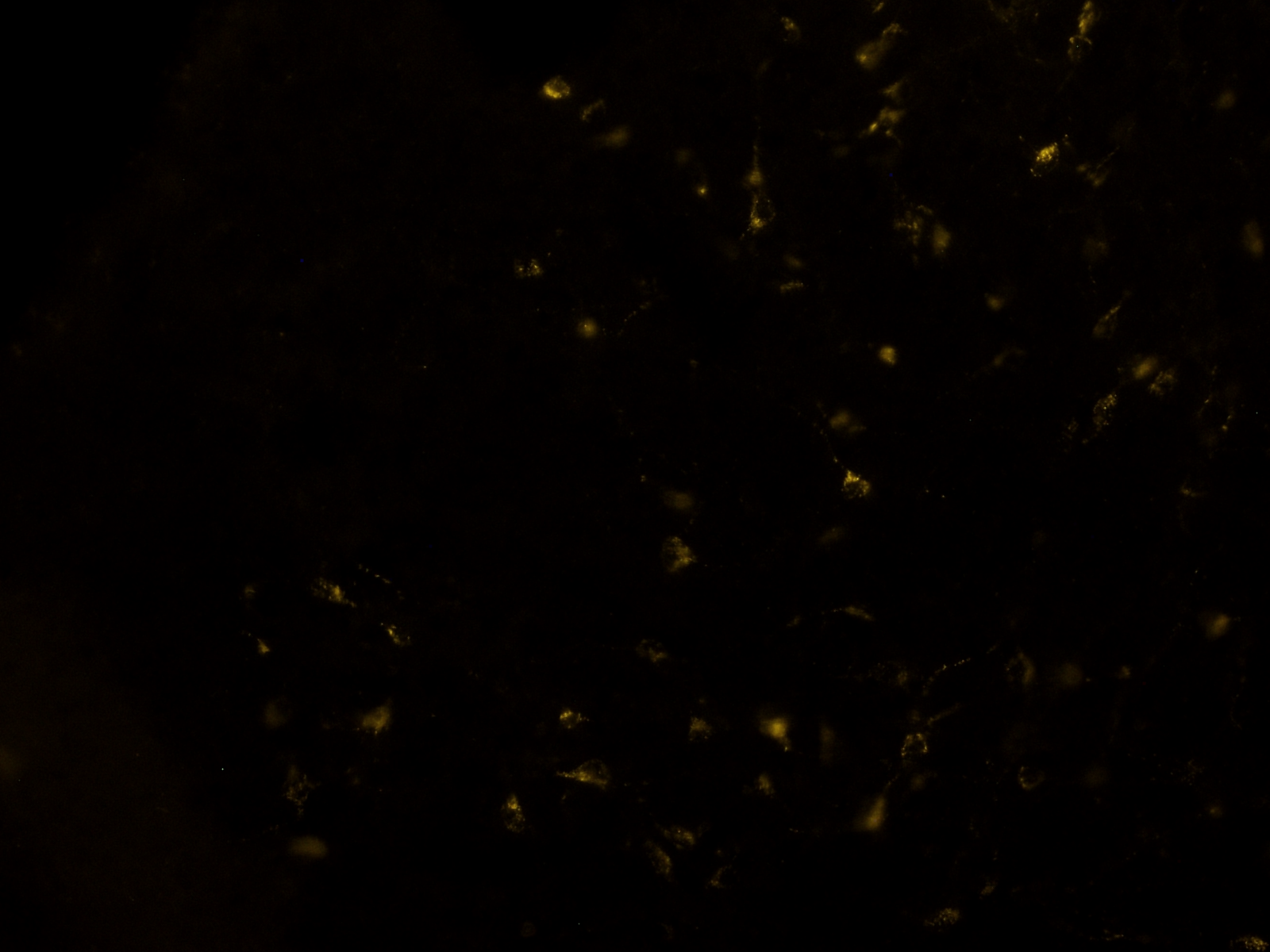}
        \caption{}
    \end{subfigure}
    \\
    \begin{subfigure}{0.45\textwidth}
        \centering
        \includegraphics[width=\linewidth]{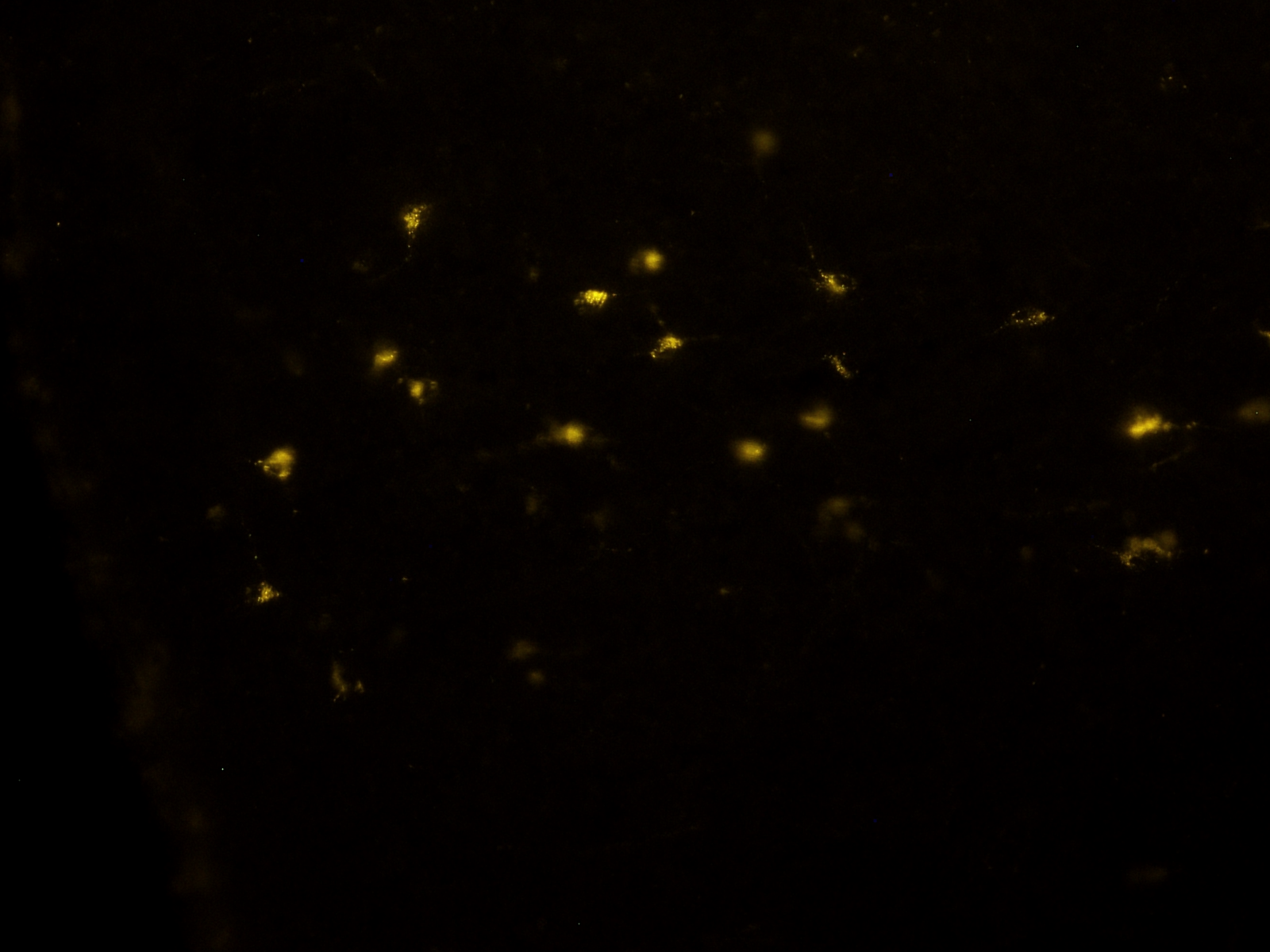}
        \caption{}
    \end{subfigure}
    \begin{subfigure}{0.45\textwidth}
        \centering
        \includegraphics[width=\linewidth]{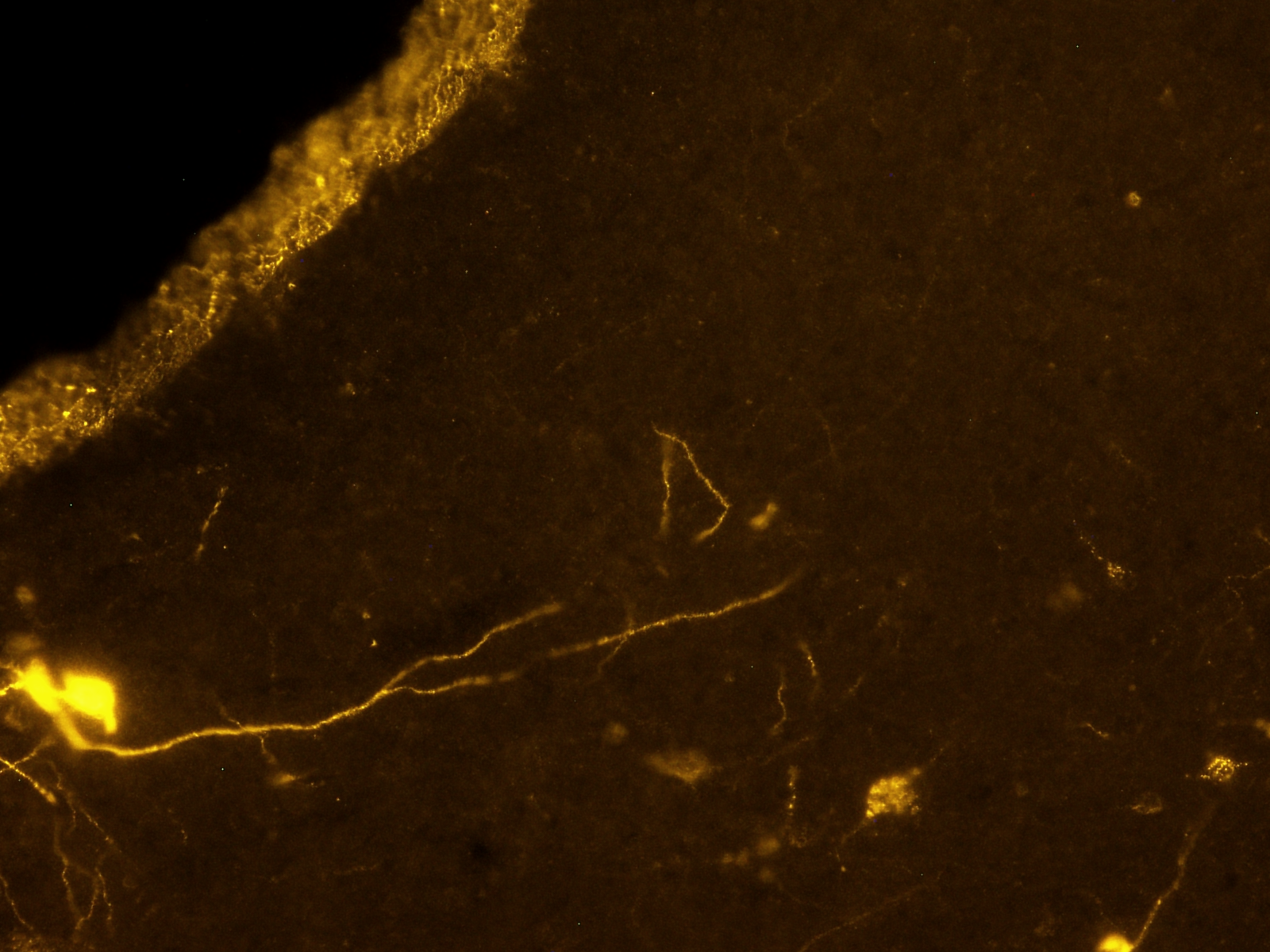}
        \caption{}
    \end{subfigure}
    
    \caption{FNV v2 - yellow sample images: these pictures depict brain tissues observed through fluorescence microscopy. Neuronal cells appear as yellow stains of different shape, size, and orientation, showing high variability in color features.}
    \label{fig:sample_images}
\end{figure}

The images in this collection depict rodent brain tissue sections that have been injected with retrograde tracers to investigate the functional relationships between brain regions during torpor onset. Neuronal cells of interest manifest as yellow fluorescent signals with heterogeneous sizes, shapes, and hue variations (see \Cref{fig:sample_images}). Following data acquisition, domain experts performed meticulous pixel-wise annotations to facilitate the development of supervised deep learning solutions.

Although the dataset comprises a moderate number of complete images (283), which would typically be insufficient for training large-scale computer vision models, the high resolution of each microscopy capture (1200 × 1600 pixels) enables an effective data amplification strategy through random cropping techniques. By extracting multiple smaller image regions from each high-resolution sample, we substantially increase the size of the adequate training dataset without compromising image quality or cellular context.

To further enhance model generalization capabilities, we implemented a comprehensive \textit{data augmentation} protocol leveraging the \texttt{fastai} library's transformation framework. Our augmentation strategy systematically applies the following transformations:

\begin{itemize}
    \item \textit{intensity normalization}: standardize pixel values into the [0,1] range, facilitating stable gradient flow during neural network training;

    \item \textit{photometric adjustments}: controlled lighting variations to simulate slight microscope illumination differences;
    
    \item \textit{geometric scaling}: random zoom transformations, enhancing model robustness to cells of varying sizes and distances from the focal plane;
    
    \item \textit{rotational transformations}: random rotations up to ±30 degrees, promoting orientation invariance in cellular detection regardless of tissue section positioning.
\end{itemize}

Notably, our implementation deliberately excludes image warping operations to preserve the authentic morphological characteristics of neuronal cells, which represent critical diagnostic features for accurate counting and classification.

This systematic data augmentation approach effectively addresses the challenge of limited dataset size in biomedical imaging applications while simultaneously enhancing model generalization across diverse experimental conditions. The transformations introduce controlled variability that corresponds to real-world imaging scenarios without compromising the fundamental relationship between input images and their corresponding segmentation maps.

The integration of these preprocessing and augmentation techniques within our standardized MLOps pipeline specifically addresses the challenges posed by fluorescent microscopy imaging while significantly improving model robustness across heterogeneous conditions. For additional details regarding data acquisition protocols and annotation methodologies, readers are referred to the original research publications \cite{clissa2024scidata, clissa2023fluocells}.

\subsection{ML Methodology}

In this work, we approached the task of cell recognition and counting using a semantic segmentation framework. This involves binary classification at the pixel level, where the model learns to distinguish background pixels (class 0) from foreground signal pixels (class 1), which correspond to the cells. Once this classification is complete, individual objects are defined as the number of such connected components determines sets of connected pixels, and the final count is determined. While alternative methods exist, this semantic segmentation approach offers advantages in both practical utility and interpretability by providing visual justification for detected objects that contribute to the final count.

To establish a controlled experimental environment suitable for demonstrating MLOps benefits, we conducted targeted ablation studies focusing exclusively on loss function variations while maintaining consistency across all other training parameters. This simplified setting enables the clear isolation of a specific component's impact, without introducing confounding variables that might obscure the MLOps workflow demonstration.

Specifically, we adopted the cell-ResUNet architecture~\cite{morelli2021cresunet} as our base model and trained it from scratch under identical conditions. Despite the extensive array of tunable hyperparameters available in deep learning systems—including learning rate schedules, batch sizes, optimization algorithms, and architectural modifications—we deliberately constrained our experimental scope to variations in the loss function alone. This simplification serves to create an accessible demonstration case that highlights MLOps benefits without overwhelming technical complexity.

The ablation study systematically evaluated two distinct loss functions: Dice~\cite{sudre2017dice} and Focal~\cite{lin2017focal},
while maintaining all other training parameters constant. This controlled experimental design enables direct comparison of how different loss functions affect model performance for the cell counting task. Furthermore, the MLOps framework described in Section 2.4 is deliberately designed with extensibility in mind, allowing for straightforward incorporation of additional hyperparameter explorations in future investigations.

\subsection{ML Training \& Testing}\label{sec:training}

To demonstrate established best practices for training machine learning models in cell counting applications, we implemented a methodologically rigorous experimental protocol that addresses specific challenges inherent in fluorescence microscopy data. Our approach incorporates several key training techniques that are particularly valuable in medical image analysis contexts where data variability and class imbalance are common obstacles.
The training phase was conducted with a 75/25 training/validation split, alternating between Dice~\cite{sudre2017dice} and Focal~\cite{lin2017focal} loss functions.

These specific loss functions were selected to address distinct challenges in the semantic segmentation task.
Dice loss provides an effective solution for handling class imbalance in segmentation tasks by directly optimizing the overlap between predicted and ground truth segments. This is particularly advantageous in our fluorescence microscopy images where cellular regions typically constitute a small fraction of the total image area.
Focal loss introduces a modulating factor that down-weights contributions from well-classified examples while emphasizing challenging cases. This characteristic is especially beneficial for addressing the varying cell morphologies and intensity distributions in our dataset, enabling the model to focus on difficult-to-classify boundary regions and cells with ambiguous features.

For each loss function, training was repeated three times with different random seeds for data splitting and weight initialization to assess statistical variability in the results. This replication strategy follows established practices in machine learning experimentation to ensure robust performance evaluation. All other hyperparameters, including batch size, learning rate, and early stopping criteria, were systematically controlled and kept unchanged across experimental conditions to isolate the effect of the loss function.

The training procedure was initialized with learning rates determined through the "learning rate test" methodology~\cite{smith2019hyperparms}, which systematically identifies optimal learning rate ranges through incremental increases. After 100 epochs of initial training, we implemented an early stopping mechanism based on validation loss performance, with a patience parameter of 50 epochs to terminate training when no further improvement was observed. This approach prevents overfitting while ensuring sufficient learning opportunities for the model to converge.

After training, we conducted a holistic model assessment focusing on segmentation, detection, and counting performance. For segmentation, true positives (TP) were defined as true and predicted objects with an overlap greater than 40\% (IoU $>$ 0.4). For detection, TP was determined based on a minimum distance between the centers of objects being less than 1.5 times the average cell diameter ($d <$ 40 pixels). Segmentation and detection performances were measured using the F1 score, a standard metric in similar applications.

Counting performance was evaluated using the \textit{mean percentage error} as the primary metric, providing a comprehensive assessment of the model's accuracy in cell quantification. For a detailed overview of the training settings and metric implementations, readers are referred to the code available at \url{https://github.com/clissa/fluocells-MLOps}.

\subsection{Continuous Integration for MLOps in Cell Counting}

Continuous Integration (CI) is an essential element in the context of an MLOps project as shown in figure \ref{fig:pipeline}. In the represented pipeline, CI enables the continuous integration of code changes, ensuring a close connection between the various stages of the Machine Learning (ML) model lifecycle, from design to operation and monitoring. Adopting CI in this context ensures that updates and improvements to the ML model for cell counting are incorporated without compromising its stability.
In the diagram, continuous integration is represented by a pipeline that includes automatic retraining and model metrics monitoring, which are crucial elements for maintaining the system's precision and reliability. Tools like $GitHub$, shown in the image, facilitate code management and version control, enabling collaboration between developers and data scientists. Additionally, integrating data from model inference with CI/CD processes ensures that every new change is automatically tested, reducing the risk of errors during the production phase.
Using CI in such projects promotes collaboration between interdisciplinary teams, increasing workflow efficiency. In sensitive contexts such as medical research and healthcare applications, implementing CI helps minimize human errors, thereby increasing the reliability of cell counting results, as highlighted by the carbon footprint evaluation ($CodeCarbon$), which supports the sustainability of the entire process.
The technologies used in the project represented in the image provide further advantages to CI in the context of cell counting:
\begin{itemize}
    \item \textbf{Jupyter Notebook}: used for training and experimenting with ML models, it allows data scientists to iterate quickly on ideas, visualize results interactively, and share their progress with other team members. This facilitates debugging and continuous validation of models during the development process.
   \item \textbf{Neptune.ai}: a tool for managing the lifecycle of Machine Learning models. Neptune.ai allows you to log and monitor model metrics, facilitating the traceability and reproducibility of results, which are essential for maintaining quality and reliability in cell counting.
    \item \textbf{Google Cloud Platform (GCP)}: used for model inference and CI/CD pipeline hosting, GCP provides scalable and resilient infrastructure. This is essential for handling the large volumes of data required for cell counting and for reliably deploying models in production.
    \item \textbf{CodeCarbon}: used to assess the carbon footprint during model training and inference, CodeCarbon provides valuable insights to optimize energy consumption in ML operations, promoting sustainability. This is particularly important in research, where energy efficiency and environmental responsibility are key values.
\end{itemize}    
These elements complete the picture of the illustrated MLOps pipeline, highlighting the integration of open source tools and cloud platforms to ensure efficiency, collaboration, and sustainability throughout the lifecycle of Machine Learning models. With these technologies, CI allows for an iterative and robust management of the cell counting model, ensuring precision and reliability at every stage of the process.

\subsection{Continuous Delivery for MLOps in Cell Counting}

In modern MLOps workflows, Continuous Delivery (CD) is crucial to ensure that machine learning models are deployed seamlessly and efficiently, making them available for production use with minimal manual intervention. CD ensures that model updates, retraining, and deployments are automated, reducing downtime and minimizing human error. This automation is particularly critical in healthcare domains, such as cell counting, where accurate and timely results are essential to diagnostics and medical research.

In our MLOps framework, we implemented continuous delivery to facilitate rapid and reliable model deployment. The experiments were initially developed and tracked using Neptune.ai, while local computations were performed on an NVIDIA T1000 GPU. However, we leveraged Google Cloud's Vertex AI for production scalability, model serving, and endpoint management.

Architecture Overview
The overall architecture of the CI/CD pipeline, specifically tailored for MLOps in cell counting, integrates several key components:
\begin{enumerate}
    \item \textbf{Data Management:}
    \begin{itemize}
        \item initial dataset storage: datasets for training the model were stored on Google Cloud Storage, ensuring they were accessible for preprocessing and model training tasks.
        \item Vertex AI pipelines were used to handle data preprocessing, feature extraction, and model training on a scalable cloud environment, ensuring that data transformations and feature engineering steps are repeatable and standardized.
    \end{itemize}
    
    \item \textbf{Model Training and Experimentation:}
    \begin{itemize}
        \item local experiments: the initial model training was conducted locally using an NVIDIA T1000 GPU to test performance and model configurations. Neptune.ai was utilized to log metrics, hyperparameters, and experiment results, enabling efficient experiment tracking and comparison.
        \item Vertex AI AutoML: after local experiments, Vertex AI was employed to scale up model training using its AutoML capabilities, allowing the team to rapidly iterate through different model architectures and configurations on larger datasets.
    \end{itemize}
    
    \item \textbf{Continuous Delivery (CD):}
    \begin{itemize}
        \item once a model meets performance criteria, it is automatically deployed through a CI/CD pipeline hosted on Google Cloud. The model is containerized using Docker and integrated with Vertex AI Endpoints for real-time API access.
        \item CD is orchestrated using GitHub Actions, which triggers a deployment workflow every time a new model is registered. The API endpoints are automatically updated to serve the most recent one, validated model, ensuring continuous availability of the latest version.
        \item the CD pipeline also integrates Vertex AI Model Monitoring, which detects data drift or performance degradation, triggering automated retraining if necessary. This ensures that the model remains robust and adaptive to evolving data.
    \end{itemize}
    
    \item \textbf{API Management and Deployment:}
    \begin{itemize}
        \item the deployed models are exposed via API endpoints managed by Vertex AI, ensuring that medical professionals and other systems can access the models with minimal latency. These APIs are secured and optimized for high throughput, supporting real-time cellcounting applications.
        \item to ensure scalability and high availability, the API is containerized in Docker. It can be deployed in any cloud or on-premises environment that supports Docker, further enhancing the system's flexibility.
        \item additional datasets and retraining: the architecture supports the addition of new datasets through Cloud Storage, allowing continuous improvement of the models through regular retraining pipelines.
    \end{itemize}
\end{enumerate}

The architecture supports the addition of new datasets through Cloud Storage, enabling continuous model improvement through regular retraining pipelines. This feature is crucial in domains such as cell counting, where new datasets can enhance the generalizability and accuracy of models, particularly for edge cases or rare cell types.

\begin{figure}[h]
    \includegraphics[width=1.0\textwidth]{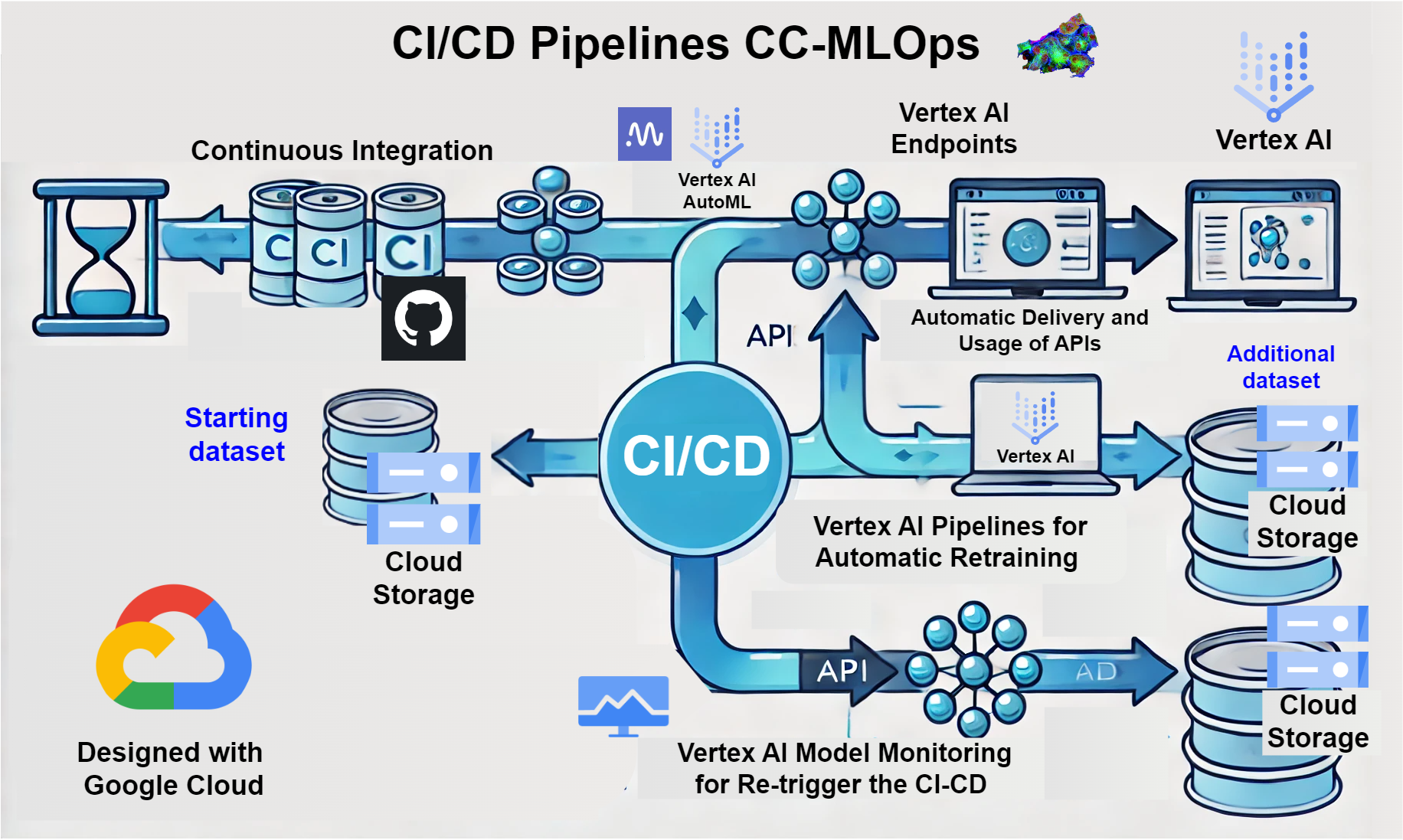}
    \caption{A visual overview of CI/CD pipelines for CC-MLOps leveraging Google Cloud and Vertex AI. It is  illustrated data ingestion, automated model training, continuous delivery, and model monitoring for retraining}
\end{figure}

The specific challenges of cell counting in medical diagnostics require an architecture that is scalable, highly reliable, and secure. The CC-MLOps (Continuous Integration and Continuous Delivery for MLOps) architecture implemented here provides several advantages:
\begin{itemize}
    \item scalability: by leveraging VertexRegulatory compliance and security: Maintaining compliance with standards such astasets and high-throughput API requests, critical in healthcare settings where real-time or near-real-time results are necessary.
    \item automated monitoring and retraining: Vertex AI Model Monitoring ensures that the system remains performant, automatically retraining models when performance metrics degrade. This is crucial for maintaining accuracy over time, especially when the underlying data distribution shifts.
    \item regulatory compliance and security: maintaining compliance with standards like the EU Medical Device Regulation (EUMDR) and the International Medical Device Regulators Forum (IMDRF) is critical in healthcare. CD enforces strict controls over API access, ensuring that only trusted entities can interact with the models. By utilizing secure API tokens and managing access through Vertex AI, the system adheres to the highest security standards.
    \item real-time availability: the deployed API allows real-time interaction with the model, enabling automated cell counting and seamless integration with clinician workflows. This real-time capability is essential for diagnostics and treatment planning applications, where timely results can impact patient outcomes.
\end{itemize}

This architecture, which combines local GPU experimentation and cloud-based production deployment, enables the best of both worlds: rapid prototyping and experimentation on local hardware, with scalable and reliable deployment in the cloud.

\subsection{Continuous Training}

Continuous Training (CT) in machine learning (ML) involves regularly updating and retraining models with new data. This practice is crucial for maintaining model prediction performance, particularly when encountering new, unseen data. Models' sensitivity to changes in real-world phenomena and user behavior necessitates periodic retraining. 

CT is essential in our case study due to the constantly evolving nature of medical data. In healthcare, new insights, procedures, and patient characteristics continually reshape the landscape, and models must be regularly updated to capture these changes. Accurate cell counting, a critical aspect of medical research and diagnostics, requires models to be sensitive to data pattern shifts.

CT addresses challenges such as data and concept drift, which occur when the data and concepts no longer accurately represent reality. In the CC-MLOps framework, the importance of CT is evident when considering changes in patient characteristics, advancements in medical practices impacting cellular properties, and external factors like economic conditions influencing health. Recognizing these potential drifts is crucial, and CT provides a mechanism to adapt the models accordingly.

For our use case, triggering events for CT in cell counting models can be defined based on changes to data, models, or code. Physicians' experience may play a crucial role in identifying triggers, enabling a strategy that aligns with their clinical expertise. Alternatively, periodic retraining, although less effective in detecting abrupt drifts, provides robustness to changes not detected by trigger-based approaches. 

CT is inherently connected with Continuous Monitoring, which defines the retraining strategy. This integration ensures that the model remains robust and accurate over time. Additionally, CT requires access to new data for model updates. When obtaining real-world data is challenging, synthetic data becomes a viable and increasingly popular option.


\subsection{Continuous Monitoring}

Continuous Monitoring (CM) involves systematically collecting, analyzing, and interpreting data to identify potential issues, anomalies, or risks, thereby ensuring the ongoing performance and security of ML models. 

In our CC-MLOps framework, CM plays a central role in safeguarding the accuracy and reliability of cell counting processes. The healthcare industry demands rigorous monitoring to identify issues such as deteriorating model performance, biases, privacy violations, cyber-attacks, or security breaches. CM is essential for ensuring patient safety and meeting the regulatory requirements and ethical standards of healthcare AI systems.

The importance of CM in the CC-MLOps framework is underscored by its ability to detect dangerous scenarios that could harm patients or healthcare organizations. By establishing effective monitoring mechanisms, CM acts as a proactive measure to identify issues before they escalate, which is crucial in medical research and diagnostics, where precision and accuracy are paramount~\cite{garg2021continuous}.

CM is implemented in our specific use case by periodically collecting and analyzing real-time model performance data. We utilize Neptune.ai as our monitoring platform, seamlessly integrating with the model API deployed on major cloud providers, including Google Cloud Platform, Amazon Web Services, and Microsoft Azure. Neptune.ai provides alerts and dashboards to verify model performance in the production environment. It is configured to monitor various metrics, including machine learning (ML) performance metrics and hardware resource usage, such as CPU, GPU, and memory. This comprehensive monitoring strategy enables accurate quantification of energy consumption in the cell counting process, while performance metrics help detect data drifts, ensuring the ongoing reliability of ML models \cite{soh2020machine}.

Additionally, CM extends to tracking API usage through Render Dashboards, enabling monitoring of incoming calls and logs and identifying potential security breaches or performance issues. This comprehensive approach to CM in the CC-MLOps framework offers valuable insights for selecting between different deep learning models, assessing their downstream impact, and ensuring the overall security and performance of the cell counting process.



\subsection{Explainability}

Deep neural networks often function as opaque algorithms, or ``black boxes'', producing predictions through complex calculations. This opacity presents a challenge in providing clear explanations for specific outcomes, especially in scientific and medical contexts where understanding the network's internal representations is crucial \cite{millerxai,byrinexai}. In the context of CC-MLOps, we aim to explain the model's predictive outcomes.

To achieve this, we used a widely adopted technique for Convolutional Neural Networks (CNNs) known as Gradient-weighted Class Activation Mapping (Grad-CAM) \cite{gradcampaper}. Grad-CAM uses the gradients of the network's layer to identify areas of focus within the input. Doing so provides insights into whether the neural model has learned to prioritize relevant aspects for accurate predictions. This information is visualized as a heatmap superimposed on the input image, highlighting the pixels where the model placed greater attention.

Grad-CAM is particularly effective with CNNs and has been applied in various applications~\cite{huff2021interpretation, PANWAR2020110190}. We have integrated Grad-CAM into our CC-MLOps pipeline to explain the cell counting process. These explanations can be accessed through API calls, allowing users to gain insights into model decisions.

\begin{figure}[t]
    \includegraphics[width=1\textwidth]{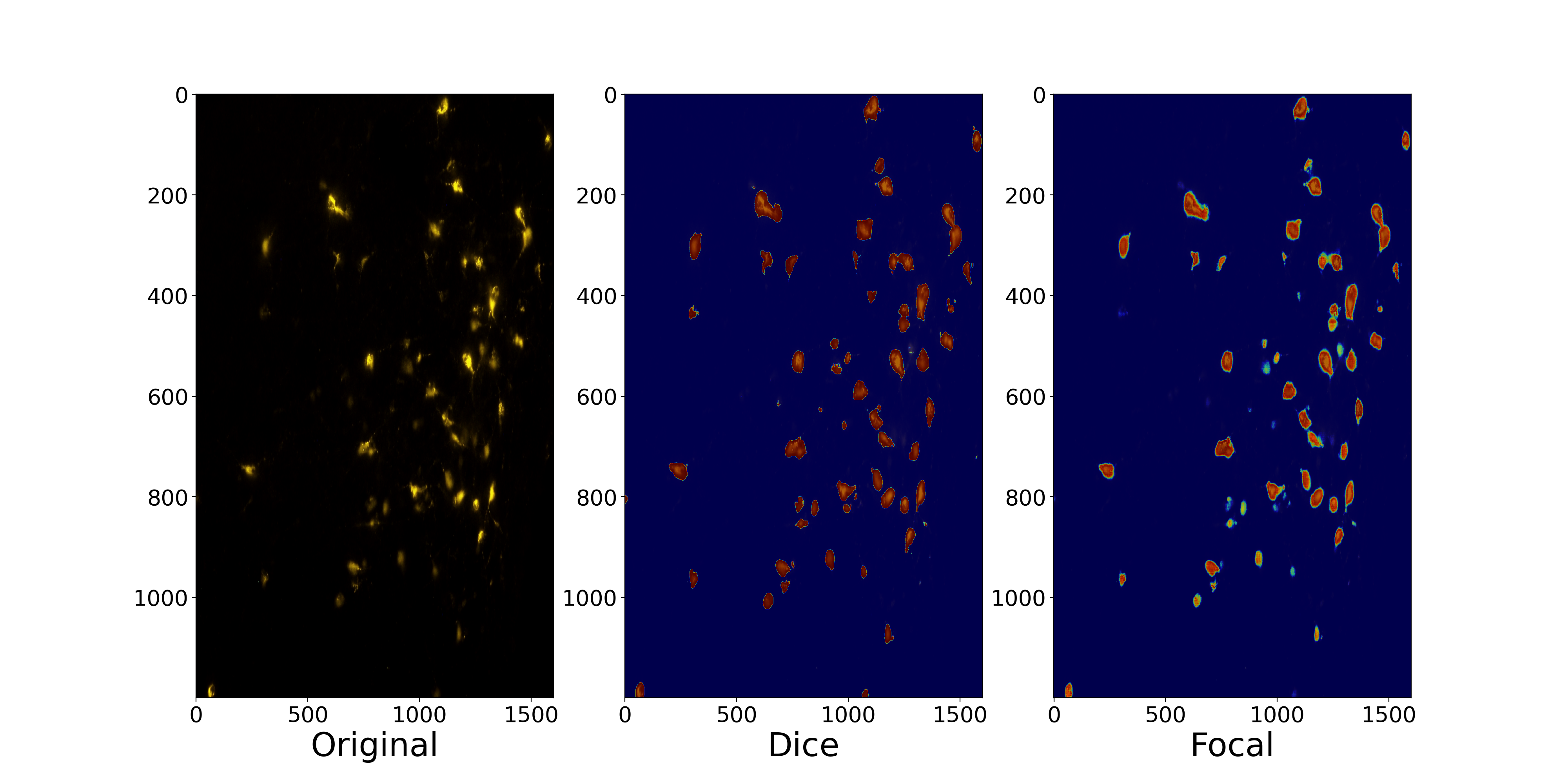}
    \caption{Grad-CAM heatmaps. From left to right, the figure depicts the original image followed by the Grad-CAM visualizations for the models trained with dice loss and focal loss, respectively. 
    }
    \label{fig:cam}
\end{figure}

In Figure \ref{fig:cam}, we display the heatmaps of the gradients in the output layer for models trained with different loss functions. Both models successfully capture the cells in the image, demonstrating the effectiveness of Grad-CAM in providing explainability. 

\subsection{Sustainability}

Training machine learning (ML) models, huge ones, involves significant energy consumption, making sustainability an increasingly important consideration in deploying AI technologies. Measuring the environmental impact of these models is crucial to identifying those that minimize carbon dioxide (CO$_2$) emissions, especially during CPU and GPU-intensive phases of the pipeline.

To address this issue, we implemented a tracking system exploiting Python's \textit{CodeCarbon} package to monitor CPU and GPU usage during training, allowing us to calculate the associated CO$_2$ emissions. 
\Cref{tab:emissions} reports the total energy consumption and emissions for the experiments conducted as described in \cref{sec:training}.
The data reveals a substantial disparity between the two loss functions examined: models trained with Dice loss consumed approximately twice the energy (0.23 KWh CPU, 0.49 KWh GPU) compared to those trained with Focal loss (0.11 KWh CPU, 0.24 KWh GPU). This differential resource utilization directly translated to corresponding carbon footprint, with Dice-based training producing 0.193 kg compared to Focal's 0.096 kg.

These findings underscore how algorithmic choices that may seem innocuous from a purely performance-oriented perspective can have significant environmental implications. This observation underscores the importance of incorporating sustainability metrics into the model selection process within the MLOps framework.

Therefore, striking a balance between energy efficiency and model accuracy is essential. Ultimately, deciding which model to deploy should align with user needs and preferences, considering both performance and environmental impact. Additionally, it is essential to acknowledge that the energy cost of training can be minimal compared to the emissions produced during inference, particularly when handling a high volume of API requests. 


\begin{table}[t]
\centering
\begin{tabular}{l|c|c|c}
\hline
\textbf{Model} & \textbf{CO$_2$ (Kg)} & \textbf{CPU (KWh)} & \textbf{GPU (KWh)} \\ \hline
Dice  & 0.193 & 0.23 & 0.49 \\ \hline
Focal & 0.096 & 0.11 & 0.24 \\ \hline
\end{tabular}
\caption{Average CO$_2$ emissions and energy consumption for model trained with Dice loss and Focal loss. 
}
\label{tab:emissions}
\end{table}

\section{Conclusion}

In this paper, we introduced CC-MLOps, a comprehensive MLOps framework for cell counting in fluorescent microscopy images. Building upon our previous work on the taxonomy and methodology of MLOps, as presented in Testi et al. (2022), we demonstrate the application of this framework to the specific challenge of counting neuronal cells in rodent brain tissues from the Yellow Collection of the Fluorescent Neuronal Cells v2 archive.
Our implementation leverages the cell-ResUNet architecture and demonstrates the effects of two alternative loss functions, Dice and Focal, to optimize semantic segmentation performance. Through a practical use case related to torpor onset studies, we showed how our framework integrates data acquisition, model training, monitoring, explainability, and sustainability considerations within a robust MLOps pipeline. The implementations of continuous integration using GitHub, experiment tracking with Neptune.ai, and cloud deployment on Google Cloud Platform's Vertex AI establish a system that streamlines workflows, reduces human error, and enables scalability.
Despite its advantages, our framework has notable limitations. The CC-MLOps framework has not been fully implemented in production environments, which limits our understanding of how it would perform under real-world conditions. As evident from our experimental setup, which utilizes an NVIDIA T1000 GPU for local computation, the current implementation primarily focuses on pre-production validation. Integration with existing laboratory systems, such as LIMS, would require additional development, particularly to accommodate the variability in fluorescent microscopy images with different cell sizes, shapes, and hues, as observed in our data set. Although our Grad-CAM implementation provides some explainability, the visualization technique could be perceived as redundant with the segmentation maps already produced by the model, suggesting additional, more sophisticated explainability approaches may be needed for clinical adoption.
Future development should address these limitations by focusing on domain-adaptation techniques for the significant variability in fluorescent microscopy images. As highlighted in our sustainability analysis, which measured CO$_2$ emissions ranging from 0.096 to 0.193 kg across different model configurations, balancing energy efficiency with model accuracy remains an important consideration for future implementations. Our continuous monitoring approach using Neptune.ai demonstrates the framework's potential; however, further development of automated triggers for retraining based on data drift would enhance its robustness. Finally, integrating the framework with the complete regulatory compliance requirements outlined in our Continuous Delivery section, particularly regarding the EU Medical Device Regulation (EUMDR) and the International Medical Device Regulators Forum (IMDRF) standards, would facilitate its adoption in clinical settings.
By addressing these challenges and pursuing these research directions, CC-MLOps can evolve from a research demonstration to a practical tool for neuronal cell counting applications in studies of torpor and other neuroscience fields, ultimately supporting the accurate quantification of cells and analysis of their activity, which is critical for advancing our understanding of cellular biology.

\section*{Acknowledgements}

The PNRR partially funded this research - M4C2 - Investimento 1.3, as part of the Partenariato Esteso PE00000013, titled ``FAIR - Future Artificial Intelligence Research'', specifically Spoke 8, ``Pervasive AI'', funded by the European Commission under the NextGenerationEU program.

\printbibliography

@inproceedings{garg2021continuous,
	title        = {{On continuous integration/continuous delivery for automated deployment of machine learning models using mlops}},
	author       = {Garg, Satvik and Pundir, Pradyumn and Rathee, Geetanjali and Gupta, PK and Garg, Somya and Ahlawat, Saransh},
	year         = 2021,
	booktitle    = {2021 IEEE fourth international conference on artificial intelligence and knowledge engineering (AIKE)},
	pages        = {25--28},
	organization = {IEEE}
}

@article{soh2020machine,
	title        = {{Machine learning operations}},
	author       = {Soh, Julian and Singh, Priyanshi and Soh, Julian and Singh, Priyanshi},
	year         = 2020,
	journal      = {Data Science Solutions on Azure: Tools and Techniques Using Databricks and MLOps},
	publisher    = {Springer},
	pages        = {259--279}
}

@misc{clissa2023fluocells,
	title        = {{Fluorescent Neuronal Cells v2}},
	author       = {Clissa, Luca and Occhinegro, Alessandra and Piscitiello, Emiliana and Taddei, Ludovico and Macaluso, Antonio and Morelli, Roberto and Squarcio, Fabio and Hitrec, Timna and Di Cristoforo, Alessia and Luppi, Marco and Amici, Roberto and Cerri, Matteo and Bastianini, Stefano and Berteotti, Chiara and Lo Martire, Viviana and Martelli, Davide and Tupone, Domenico and Zoccoli, Giovanna},
	year         = 2023,
	publisher    = {University of Bologna},
	howpublished = {\emph{AMS Acta} \url{https://doi.org/10.6092/unibo/amsacta/7347}}
}

@article{clissa2024scidata,
	title        = {{Fluorescent Neuronal Cells v2: multi-task, multi-format annotations for deep learning in microscopy}},
	author       = {Clissa, Luca and Macaluso, Antonio and Morelli, Roberto and Occhinegro Alessandra and Piscitiello Emiliana and Taddei, Ludovico and Luppi, Marco and Amici, Roberto and Cerri, Matteo and Hitrec, Timna and Rinaldi, Lorenzo and Zoccoli, Antonio},
	year         = 2024,
	journal      = {Scientific Data},
	volume       = {},
	number       = {},
	pages        = {}
}

@article{singh2024systematic,
  title={A Systematic Survey on Biological Cell Image Segmentation and Cell Counting Techniques in Microscopic Images Using Machine Learning},
  author={Singh, Harjeet and Kaur, Harpreet},
  journal={Wireless Personal Communications},
  volume={137},
  number={2},
  pages={813--851},
  year={2024},
  publisher={Springer}
}

@article{bhattarai2024deep,
  title={A Deep Learning-Based Segmentation of Cells and Analysis (DL-SCAN)},
  author={Bhattarai, Alok and Meyer, Jan and Petersilie, Laura and Shah, Syed I and Rose, Christine R and Ullah, Ghanim},
  journal={bioRxiv},
  pages={2024--05},
  year={2024},
  publisher={Cold Spring Harbor Laboratory}
}

@inproceedings{clissa2024iciap,
	title        = {{Optimizing Deep Learning Models for Cell Recognition in Fluorescence Microscopy: The Impact of Loss Functions on Performance and Generalization}},
	author       = {Clissa, Luca and Macaluso, Antonio and Zoccoli, Antonio},
	year         = 2024,
	booktitle    = {Image Analysis and Processing - ICIAP 2023 Workshops},
	publisher    = {Springer Nature Switzerland},
	address      = {Cham},
	pages        = {179--190},
	isbn         = {978-3-031-51023-6},
	editor       = {Foresti, Gian Luca and Fusiello, Andrea and Hancock, Edwin}
}

@article{morelli2021cresunet,
	title        = {{Automating cell counting in fluorescent microscopy through deep learning with c-ResUnet}},
	author       = {Morelli, Roberto and Clissa, Luca and Amici, Roberto and Cerri, Matteo and Hitrec, Timna and Luppi, Marco and Rinaldi, Lorenzo and Squarcio, Fabio and Zoccoli, Antonio},
	year         = 2021,
	journal      = {Scientific Reports},
	volume       = 11,
	number       = 1,
	pages        = 22920
}

@article{zeng2019ric_unet,
	title        = {{RIC-Unet: An improved neural network based on Unet for nuclei segmentation in histology images}},
	author       = {Zeng, Zitao and Xie, Weihao and Zhang, Yunzhe and Lu, Yao},
	year         = 2019,
	journal      = {Ieee Access},
	publisher    = {IEEE},
	volume       = 7,
	pages        = {21420--21428}
}

@article{luppi1,
	title        = {{C-Fos expression in preoptic nuclei as a marker of sleep rebound in the rat}},
	author       = {Dentico, Daniela and Amici, Roberto and Baracchi, Francesca and Cerri, Matteo and {Del Sindaco}, Elide and Luppi, Marco and Martelli, Davide and Perez, Emanuele and Zamboni, Giovanni},
	year         = 2009,
	journal      = {European Journal of Neuroscience},
	volume       = 30,
	number       = 4,
	pages        = {651--661},
	issn         = {0953816X},
	pmid         = 19686475
}

@article{luppi2,
	title        = {{Phosphorylated Tau protein in the myenteric plexus of the ileum and colon of normothermic rats and during synthetic torpor}},
	author       = {Gillis, Richard and Adams, Gary and Besong, David and Machova, Eva and Ebringerova, Anna and Harding, Stephen and Patel, Trushar},
	year         = 2016,
	journal      = {European Biophysics Journal},
	issn         = {0175-7571}
}

@article{luppi3,
	title        = {{c-Fos expression in the limbic thalamus following thermoregulatory and wake–sleep changes in the rat}},
	author       = {Luppi, Marco and Cerri, Matteo and {Di Cristoforo}, Alessia and Hitrec, Timna and Dentico, Daniela and {Del Vecchio}, Flavia and Martelli, Davide and Perez, Emanuele and Tupone, Domenico and Zamboni, Giovanni and Amici, Roberto},
	year         = 2019,
	journal      = {Experimental Brain Research},
	publisher    = {Springer Berlin Heidelberg},
	volume       = 237,
	number       = 6,
	pages        = {1397--1407},
	isbn         = {0123456789},
	issn         = 14321106,
	pmid         = 30887077
}

@article{hitrec2021reversible,
	title        = {{Reversible Tau phosphorylation induced by synthetic torpor in the spinal cord of the rat}},
	author       = {Hitrec, Timna and Squarcio, Fabio and Cerri, Matteo and Martelli, Davide and Occhinegro, Alessandra and Piscitiello, Emiliana and Tupone, Domenico and Amici, Roberto and Luppi, Marco},
	year         = 2021,
	journal      = {Frontiers in neuroanatomy},
	publisher    = {Frontiers},
	volume       = 15,
	pages        = 3
}

@article{awasthi2023fluorescence,
	title        = {{Fluorescence microscopic approach for detection of two different modes of breast cancer cell death induced by nanosecond pulsed electric field}},
	author       = {Awasthi, Kamlesh and Li, Si-Pei and Zhu, Chao-Yuan and Hsu, Hsin-Yun and Ohta, Nobuhiro},
	year         = 2023,
	journal      = {Sensors and Actuators B: Chemical},
	publisher    = {Elsevier},
	volume       = 378,
	pages        = 133199
}

@article{derevtsova2021applying,
	title        = {{Applying the expansion microscopy method in neurobiology}},
	author       = {Derevtsova, KZ and Pchitskaya, EI and Rakovskaya, AV and Bezprozvanny, IB},
	year         = 2021,
	journal      = {Journal of Evolutionary Biochemistry and Physiology},
	publisher    = {Springer},
	volume       = 57,
	number       = 3,
	pages        = {681--693}
}

@article{lee2020deep,
	title        = {{Deep-learning-based three-dimensional label-free tracking and analysis of immunological synapses of CAR-T cells}},
	author       = {Lee, Moosung and Lee, Young-Ho and Song, Jinyeop and Kim, Geon and Jo, YoungJu and Min, HyunSeok and Kim, Chan Hyuk and Park, YongKeun},
	year         = 2020,
	journal      = {Elife},
	publisher    = {eLife Sciences Publications, Ltd},
	volume       = 9,
	pages        = {e49023}
}

@article{bouma2012induction,
	title        = {{Induction of torpor: mimicking natural metabolic suppression for biomedical applications}},
	author       = {Bouma, Hjalmar R and Verhaag, Esther M and Otis, Jessica P and Heldmaier, Gerhard and Swoap, Steven J and Strijkstra, Arjen M and Henning, Robert H and Carey, Hannah V},
	year         = 2012,
	journal      = {Journal of cellular physiology},
	publisher    = {Wiley Online Library},
	volume       = 227,
	number       = 4,
	pages        = {1285--1290}
}

@article{alam2012hypothermia,
	title        = {{Hypothermia and hemostasis in severe trauma: a new crossroads workshop report}},
	author       = {Alam, Hasan B and Pusateri, Anthony E and Kindzelski, Andrei and Egan, Debra and Hoots, Keith and Andrews, Matthew T and Rhee, Peter and Tisherman, Samuel and Mann, Kenneth and Vostal, Jaroslav and others},
	year         = 2012,
	journal      = {Journal of Trauma and Acute Care Surgery},
	publisher    = {LWW},
	volume       = 73,
	number       = 4,
	pages        = {809--817}
}

@article{CERRI2021cool,
	title        = {{Be cool to be far: Exploiting hibernation for space exploration}},
	author       = {Matteo Cerri and Timna Hitrec and Marco Luppi and Roberto Amici},
	year         = 2021,
	journal      = {Neuroscience \& Biobehavioral Reviews},
	volume       = 128,
	pages        = {218--232},
	issn         = {0149-7634}
}

@article{puspitasari2021hibernation,
	title        = {{Hibernation as a Tool for Radiation Protection in Space Exploration}},
	author       = {Puspitasari, Anggraeini and Cerri, Matteo and Takahashi, Akihisa and Yoshida, Yukari and Hanamura, Kenji and Tinganelli, Walter},
	year         = 2021,
	journal      = {Life},
	publisher    = {Multidisciplinary Digital Publishing Institute},
	volume       = 11,
	number       = 1,
	pages        = 54
}

@inproceedings{sudre2017dice,
	title        = {{Generalised dice overlap as a deep learning loss function for highly unbalanced segmentations}},
	author       = {Sudre, Carole H and Li, Wenqi and Vercauteren, Tom and Ourselin, Sebastien and Jorge Cardoso, M},
	year         = 2017,
	booktitle    = {Deep Learning in Medical Image Analysis and Multimodal Learning for Clinical Decision Support },
	publisher    = {Springer International Publishing},
	pages        = {240--248}
}

@inproceedings{lin2017focal,
	title        = {{Focal loss for dense object detection}},
	author       = {Lin, Tsung-Yi and Goyal, Priya and Girshick, Ross and He, Kaiming and Doll{\'a}r, Piotr},
	year         = 2017,
	booktitle    = {Proceedings of the IEEE international conference on computer vision},
	pages        = {2980--2988}
}

@article{smith2019hyperparms,
	title        = {{A disciplined approach to neural network hyper-parameters: Part 1--learning rate, batch size, momentum, and weight decay}},
	author       = {Smith, Leslie N},
	year         = 2018,
	journal      = {arXiv preprint arXiv:1803.09820}
}

@article{poon2023dataset,
	title        = {{A dataset of rodent cerebrovasculature from in vivo multiphoton fluorescence microscopy imaging}},
	author       = {Poon, Charissa and Teikari, Petteri and Rachmadi, Muhammad Febrian and Skibbe, Henrik and Hynynen, Kullervo},
	year         = 2023,
	journal      = {Scientific Data},
	publisher    = {Nature Publishing Group UK London},
	volume       = 10,
	number       = 1,
	pages        = 141
}

@article{PANWAR2020110190,
	title        = {{A deep learning and grad-CAM based color visualization approach for fast detection of COVID-19 cases using chest X-ray and CT-Scan images}},
	author       = {Harsh Panwar and P.K. Gupta and Mohammad Khubeb Siddiqui and Ruben Morales-Menendez and Prakhar Bhardwaj and Vaishnavi Singh},
	year         = 2020,
	journal      = {Chaos, Solitons \& Fractals},
	volume       = 140,
	pages        = 110190,
	issn         = {0960-0779}
}

@article{huff2021interpretation,
	title        = {{Interpretation and visualization techniques for deep learning models in medical imaging}},
	author       = {Huff, Daniel T and Weisman, Amy J and Jeraj, Robert},
	year         = 2021,
	journal      = {Physics in Medicine \& Biology},
	publisher    = {IOP Publishing},
	volume       = 66,
	number       = 4,
	pages        = {04TR01}
}

@article{grishaginautomatcellcounting,
	title        = {{Automatic cell counting with ImageJ}},
	author       = {Ivan V. Grishagin},
	year         = 2014,
	journal      = {Analytical Biochemistry},
	publisher    = {Elsevier},
        volume       = 473,
	number       = {},     
	pages        = {63-65}
}

@article{janapareddicloud,
	title        = {Edge Impulse: an MLOps platform for tiny machine learning},
	author       = {Janapa,Reddi and Shawn, Hymel and Colby,Banbury and Daniel, Situnayake and Alex, Elium and Carl, Ward and Mat, Kelcey and Mathijs, Baaijens and Mateusz, Majchrzycki and Jenny,Plunkett and David, Tischler and Alessandro,Grande and Louis,Moreau and Dmitry, Maslov and Artie, Beavis and Jan Jongboom, Vijay},
	year         = {2023},
	journal      = {MLSys Proceedings},
	volume       = {},    
	number       = {},  
	pages        = {}   
}

@article{cobparroagronomy,
	title        = {Fostering Agricultural Transformation through AI: An Open-Source AI Architecture Exploiting the MLOps Paradigm},
	author       = {Antonio Carlos, Cob-Parro and Yerhard, Lalangui and Yerhard,Lalangui and Raquel, Lazcano },
	year         = {2024},
	journal      = {Agronomy},
	volume       = {},    
	number       = {},  
	pages        = {}   
}

@article{lombardotwins,
	title        = {Digital Twin for Continual Learning in Location Based Services},
	author       = {Gianfranco,Lombardo and 
Marco,Picone and Marco,Mamei and Monica,Mordonini andAgostino,Poggi},
	year         = {2024},
	journal      = {Engineering Application of Artifical Intelligence},
	volume       = {127},    
	number       = {A},  
        publisher    = {Elsevier},
	pages        = {}   
}

@article{testimlops,
	title        = {MLOps: A Taxonomy and a Methodology},
	author       = {Matteo,Testi and 
Matteo,Ballabio and Emanuele,Frontoni and Goilio,Iannello and Sara,Moccia and Paolo,Soda and Gennaro,Vessio},
	year         = {2022},
	journal      = {},
	volume       = {},    
	number       = {},  
        publisher    = {IEEE Access},
	pages        = {}   
}

@article{crispdmprocess,
	title        = {A Systematic Literature Review on Applying CRISP-DM Process Model},
	author       = {Cristopher,Schröer and 
Felix,Kruse and Jorge Marx, Gómez},
	year         = {2021},
	journal      = {Procedia Computer Science},
	volume       = {181},    
	number       = {},  
        publisher    = {Elsevier},
	pages        = {526-534}   
}

@article{millerxai,
	title        = {Explanation in Artificial Intelligence: Insights from the social science},
	author       = {Tim, Miller},
	year         = {2019},
	journal      = {Artificial Intelligence},
	volume       = {267},    
	number       = {},  
        publisher    = {elsevier},
	pages        = {1-38}   
}

@article{byrinexai,
	title        = {Counterfactuals in Explainable Artificial Intelligence (XAI): Evidence from Human Reasoning},
	author       = {Ruth M. J.,Byrne},
	year         = {2019},
	journal      = {},
	volume       = {},    
	number       = {},  
        publisher    = {Twenty-Eighth International Joint Conference on Artificial Intelligence},
	pages        = {}   
}

@article{gradcampaper,
	title        = {Grad-CAM: Visual Explanations from Deep Networks via Gradient-based Localization},
	author       = {Ramprasaath R., Selvaraju and Michael, Cogswell and Abhishek, Das and Ramakrishna, Vedantam and Devi,Parikh and  Dhruv,Batra},
	year         = {2016},
	journal      = {International Journal of Computer Vision},
	volume       = {},    
	number       = {},  
        publisher    = {Springer},
	pages        = {}   
}

@article{aicellcounter,
	title        = {AICellCounter: A Machine Learning-Based Automated Cell Counting Tool Requiring Only One Image for Training},
    author       = {Junnan Xu and Andong Wang and Yunfeng Wang and Jingting Li and Ruxia Xu and Hao Shi and Xiaowen Li and Yu Liang and Jianming Yang and Tian-Ming Gao},
	year         = {2022},
	journal      = {Neuroscience Bulletin},
	volume       = {},    
	number       = {},  
    publisher    = {Springer},
	pages        = {}   
}

@article{mizuno,
	title = {Noninvasive total counting of cultured cells using a home-use scanner with a pattern sheet},
	author       = {Mitsuru Mizuno and Yoshitaka Maeda and Sho Sanami and Takahisa Matsuzaki and Hiroshi Y. Yoshikawa and Nobutake Ozeki and Hideyuki Koga and Ichiro Sekiya},
	year         = {2024},
	journal      = {},
	volume       = {},    
	number       = {},  
        publisher    = {iScience: Cell press},
	pages        = {}   
}

@article{molder,
	title = {Non-invasive, label-free cell counting and quantitative analysis of adherent cells using digital holography},
	author       = {A. Mölder and M. Sebesta and  M. Gustafsson and L. Gisselson and A. Gjörloff Wingren and K. Alm},
	year         = {2024},
	journal      = {Journal of Microscopy},
	volume       = {},    
	number       = {},  
    publisher    = {iScience: Cell press},
	pages        = {}   
}


\end{document}